\documentclass[traditabstract]{aa} 		

\usepackage{graphicx}
\usepackage{txfonts}
\usepackage{natbib}
\usepackage{enumitem}
\usepackage{lscape}

\begin{document}

\title{Molecular gas mass functions of normal star forming galaxies since $z\sim 3$
\thanks{\textit{Herschel} is an ESA space observatory with science instruments provided by 
European-led Principal Investigator consortia and with important participation from NASA.}}

\author{S. Berta\inst{1}
        \and
	D. Lutz\inst{1}
	\and
	R. Nordon\inst{2}
	\and
        R. Genzel\inst{1}
        \and 
        B. Magnelli\inst{3}
        \and
	P. Popesso\inst{1}
	\and
	D. Rosario\inst{1}
	\and
        A. Saintonge\inst{1}
        \and
	S. Wuyts\inst{1}
        \and
	L.~J. Tacconi\inst{1}
	}

\offprints{Stefano Berta, \email{berta@mpe.mpg.de}}

\institute{Max-Planck-Institut f\"{u}r extraterrestrische Physik (MPE),
Postfach 1312, 85741 Garching, Germany.
\and
School of Physics and Astronomy, The Raymond and Beverly Sackler
Faculty of Exact Sciences, Tel-Aviv University, Tel-Aviv 69978, Israel.
\and 
Argelander-Institut f\"ur Astronomie, Universit\"at Bonn, Auf dem H\"ugel 71, 
D-53121 Bonn, Germany
}

\date{Received ...; accepted ...}

\abstract{
We use deep far-infrared data from the PEP/GOODS-\textit{Herschel} surveys 
and rest frame 
ultraviolet photometry to study the evolution of the molecular gas mass function of 
normal star forming galaxies. Computing the molecular gas mass, $M_{\rm mol}$, 
by scaling star formation rates (SFR) through depletion timescales, or combining IR luminosity 
and obscuration properties as in Nordon et al., we obtain $M_{\rm mol}$ for roughly 
700, $z=0.2-3.0$ galaxies near the star forming ``main 
sequence''. The number density of galaxies follows a Schechter function of 
$M_{\rm mol}$. The characteristic mass $M^\ast$ is found to 
strongly evolve up to $z\sim 1$, and then to flatten at earlier epochs, 
resembling the infrared luminosity evolution of similar objects. 
At z$\sim$1, our result is supported 
by an estimate based on the stellar mass function 
of star forming galaxies and gas fraction scalings from the PHIBSS survey. 
We compare our measurements to results from current models, 
finding better agreement with those that are treating star formation laws 
directly rather than in post-processing. Integrating the mass function, we 
study the evolution of the $M_{\rm mol}$ density and its density parameter
$\Omega_{\rm mol}$. 
}

\keywords{Galaxies: mass function -- Galaxies: statistics -- Galaxies: evolution -- Galaxies: star formation -- Infrared: galaxies}

\maketitle


\section{Introduction}\label{sect:intro}

Stars form in cold, dense molecular clouds, and the molecular gas content of
galaxies is an important constraint to galaxy evolution. In the 
interplay between accretion of gas, star formation, metal enrichment,
and outflows, the molecular gas content reflects the prevailing
physical processes \citep[e.g.][]{bouche2010,dave2010,lilly2013}.    
For the cosmic baryon budget \citep[e.g.][]{fukugita1998}, molecular gas 
can be of increasing relative importance at high redshifts where galaxies 
are gas-rich.

Molecular gas is generally quantified using the CO molecule as a tracer. 
Local samples include several hundred targets 
\citep[e.g.][and references therein]{saintonge2011a}, but
CO detections of star-forming, intermediate/high redshift galaxies 
are still limited to modest statistics of mostly very luminous galaxies
\citep[][and references therein]{carilli2013}. For normal star forming
galaxies, the PHIBSS survey \citep{tacconi2012} derives scaling relations on 
the basis of CO detections 
of 52 normal star forming galaxies at $z\sim1.2$ and $z\sim2.2$.
In galaxies near the star-forming ``main sequence'' 
\citep[MS, e.g.][]{noeske2007,elbaz2007,daddi2007a}, 
the molecular gas mass, $M_{\rm mol}$, scales as SFR through a depletion 
time scale $\tau_{\rm dep}$ that is only very slowly dependent on redshift
\citep{tacconi2012}. 

\citet{keres2003} reported the first attempt to derive the local molecular gas 
mass function, based on a sample of IRAS selected objects. 
To date, no derivation of a $z>0$ mass function has been possible, 
due to the heterogeneity and paucity of the available detections.

These difficulties have sparked interest in methods using dust mass as a 
tool to derive masses of the cold ISM of (distant) galaxies 
\citep[e.g.][]{magdis2011,magdis2012,scoville2012}. Dust mass is converted 
into gas mass by adopting a metallicity, and a scaling of gas-to-dust ratio
with metallicity. Typically, to derive dust masses, these methods require 
photometry on the restframe-submm tail of the SED, or accurate multi-band 
data
closer to the rest far-infrared (FIR) SED peak. Such requirements are still 
not met individually for large samples of normal high-z galaxies 
(Berta et al. in prep.).
In a study of the restframe ultraviolet (UV) and FIR 
properties of \textit{Herschel} 
galaxies, \citet{nordon2013} established a method to derive $M_{\rm mol}$
of MS galaxies on the basis of their rest-FIR luminosity and rest-UV 
obscuration: properties which are readily available for larger samples.
Here we apply the $\tau_{\rm dep}$ scaling and the Nordon et al. recipe
to the deepest  
\textit{Herschel}  \citep{pilbratt2010} FIR extragalactic 
observations and related restframe UV data to derive $M_{\rm mol}$ 
of MS galaxies, which are known to
power $\sim$90\% of the cosmic star formation \citep{rodighiero2011}.
On this basis, we construct their molecular gas mass function and study 
its evolution since $z\sim3$.

We adopt a $\Lambda$CDM cosmology 
with $\left(h,\Omega_m,\Omega_\Lambda\right) = \left(0.70, 0.27, 0.73\right)$
and a \citet{chabrier2003} initial mass function.



\section{Derivation of molecular gas mass and source selection}

We use the deepest PACS \citep{poglitsch2010} far-infrared survey,
combining the  PACS Evolutionary Probe \citep[PEP,][]{lutz2011} and 
GOODS-\textit{Herschel} \citep{elbaz2011} 
data in the GOODS-N and GOODS-S fields. We defer to \citet{magnelli2013b} 
for details about data reduction and catalog extraction.

Ancillary data come from the catalogs built by \citet{berta2011} and 
\citet[][MUSIC]{grazian2006}, both including photometry from the $U$ 
to \textit{Spitzer} IRAC bands. To these we add \textit{Spitzer} 
MIPS 24 $\mu$m \citep{magnelli2011}, IRS 16 $\mu$m data \citep{teplitz2011},
GALEX DR6 data, and a collection of optical spectroscopic redshifts \citep[see][for details]{berta2011}.  
When needed, photometric redshifts are used, reaching accuracies of $\Delta(z)/(1+z)$=0.04 and 0.06 in the two fields, respectively 
\citep{berta2011,santini2009}.

Based on current CO observations of local and $z\sim1$ objects
\citep[e.g.][]{saintonge2011a,saintonge2012,tacconi2012}, the relation 
between $M_{\rm mol}$ and SFR of MS galaxies can be described as a simple scaling with 
a depletion time scale mildly dependent on redshift: 
$M_{\rm mol}/\textrm{SFR}=\tau_{\rm dep}$, 
with $\tau_{\rm dep}=1.5\, 10^9\times(1+z)^{-1}$ $[$Gyr$]$.
Second order effects, likely driven by the actual distribution of dust and gas 
in galaxies, have been studied by \citet{nordon2013}, while investigating the   
infrared excess (IRX) and the observed UV slope $\beta$
of $1.0<z<2.5$ PEP galaxies \citep{meurer1999}.
Using a smaller sample of $z\ge 1$ galaxies with CO gas masses, 
Nordon et al. study
the link between $A_{\rm IRX}$ and molecular gas content, finding 
a tight relation between the scatter in
the $A_{\rm IRX}$-$\beta$ plane and the specific attenuation contributed 
by the molecular gas mass per young star. 
Here\footnote{$\textrm{SFR}_{\rm IR}$ is based on bolometric IR luminosity; $\textrm{SFR}_{\rm UV}$
is derived from UV continuum, uncorrected for dust attenuation.} $A_{\rm IRX}=2.5 \log ( \textrm{SFR}_{\rm IR}/\textrm{SFR}_{\rm UV}+1)$ 
represents the effective UV attenuation.
These authors derive the equation
\begin{equation}\label{eq:SA_method}
\log\left(\frac{M_{\rm mol}}{[10^9 \textrm{M}_\odot]}\right)= \log \left( \frac{A_{\rm IRX} \cdot \textrm{SFR}}{[\textrm{M}_\odot\textrm{yr}^{-1}]}\right)-0.20\left(A_{\rm IRX}-1.26\beta\right)+0.09\textrm{,}
\end{equation}
where the SFR includes both IR and UV contributions.
This provides an estimate of molecular gas mass (already including the
contribution of Helium) using only widely available integrated
rest-frame UV and FIR photometry, consistent with $z\sim 1$ depletion times by calibration. 
We use the FIR and UV photometry to apply the \citet{tacconi2012} $\tau_{\rm dep}$ scaling and the 
\citet{nordon2013} recipe to $\sim 700$ MS galaxies in our survey.
We stress again that both methods rely on calibrations based on CO observations 
and are thus valid under the assumption that the adopted values of the CO-to-molecular gas mass 
conversion factor, $\alpha_{\rm CO}$, are correct
\citep[see][]{tacconi2012,nordon2013}.

SFR$_{\rm IR, UV}$ are computed using the \citet{kennicutt1998} calibrations. Infrared luminosities, 
$L(\textrm{IR})$, are derived by SED fitting with the \citet{berta2013} templates library. Alternative methods \citep[e.g.][]{wuyts2011a,nordon2012} 
lead to equivalent results within a 10-20\% scatter \citep[see][]{berta2013}.
UV parameters are derived from the restframe 1600 and 2800 \AA\ luminosities, following the 
procedure of \citet{nordon2013}.

Uncertainties on $M_{\rm mol}$ are computed combining the systematic errors 
embedded in 
the adopted equations and statistical uncertainties due to the 
scatter on observed quantities.
\citet{tacconi2012} estimate a systematic uncertainty of 50\% in their 
$M_{\rm mol}$, which is thus reflected into the our $\tau_{\rm dep}$-based 
estimate.
\citet{nordon2013} evaluate that the accuracy of Eq. \ref{eq:SA_method} is 
0.12 dex for their collection of $z\ge 1$ MS galaxies and 0.16 dex when 
validating on $z\sim 0$ MS galaxies from \citet{saintonge2011a,saintonge2012}.
We performed 10000 Monte Carlo realizations of $M_{\rm mol}$ for each galaxy
assuming Gaussian distributions for the mentioned sources of errors
and accounting also for the contribution of 
the template library intrinsic scatter to the $L(\textrm{IR})$ uncertainty.
As a result, the median uncertainty on $M_{\rm mol}$ obtained through 
Eq. \ref{eq:SA_method} is $\sim$40\%, without accounting for 
\citet{tacconi2012} systematics, and is larger than 50\% for only $\sim$10\% of 
the sample.

Sample selection is driven by the need to compute total SFRs and by the requirements imposed by Eq. \ref{eq:SA_method}.
We limit the analysis to galaxies lying within $\left|\Delta\log(\textrm{SFR})_{\rm MS}\right|\le0.5$  
from the MS. The MS is assumed to have unit slope in the stellar mass vs. star formation rate ($M_\star$-SFR) plane,
 and a specific-SFR normalization 
varying as $\textrm{sSFR}_{\rm MS} [\textrm{Gyr}^{-1}] = 26 \times t_{\rm cosmic}^{-2.2}$ \citep{elbaz2011}. 

We apply a 3 $\sigma$ flux cut at 160 $\mu$m, the band 
that best correlates with $L(\textrm{IR})$ \citep{elbaz2011,nordon2012}.
When deriving UV parameters, it is important to avoid contamination 
of observed bands by the 2100 \AA\ carbonaceous absorption feature. Combining GALEX and optical 
photometry, we define four redshift windows: 
$0.2<z\le0.6$, $0.7<z\le1.0$, $1.0<z\le2.0$, 
and $2.0<z\le3.0$. This choice reduces the loss of sources due to restframe UV 
requirements to $<$2\%.
%
The total number of sources in each redshift bin is included in Table \ref{tab:mf2}.
Our sample contains 43 galaxies hosting an X-ray AGN, of which only 4 are Type-1.
Distributions of $L(\textrm{IR})$ and $\beta$ for the AGN galaxies are similar to those of inactive ones.
We have verified that our conclusions below are not affected by the inclusion of these AGN hosts.


\section{The molecular gas mass function}

\begin{figure*}[!ht]
\centering
\includegraphics[width=0.42\textwidth]{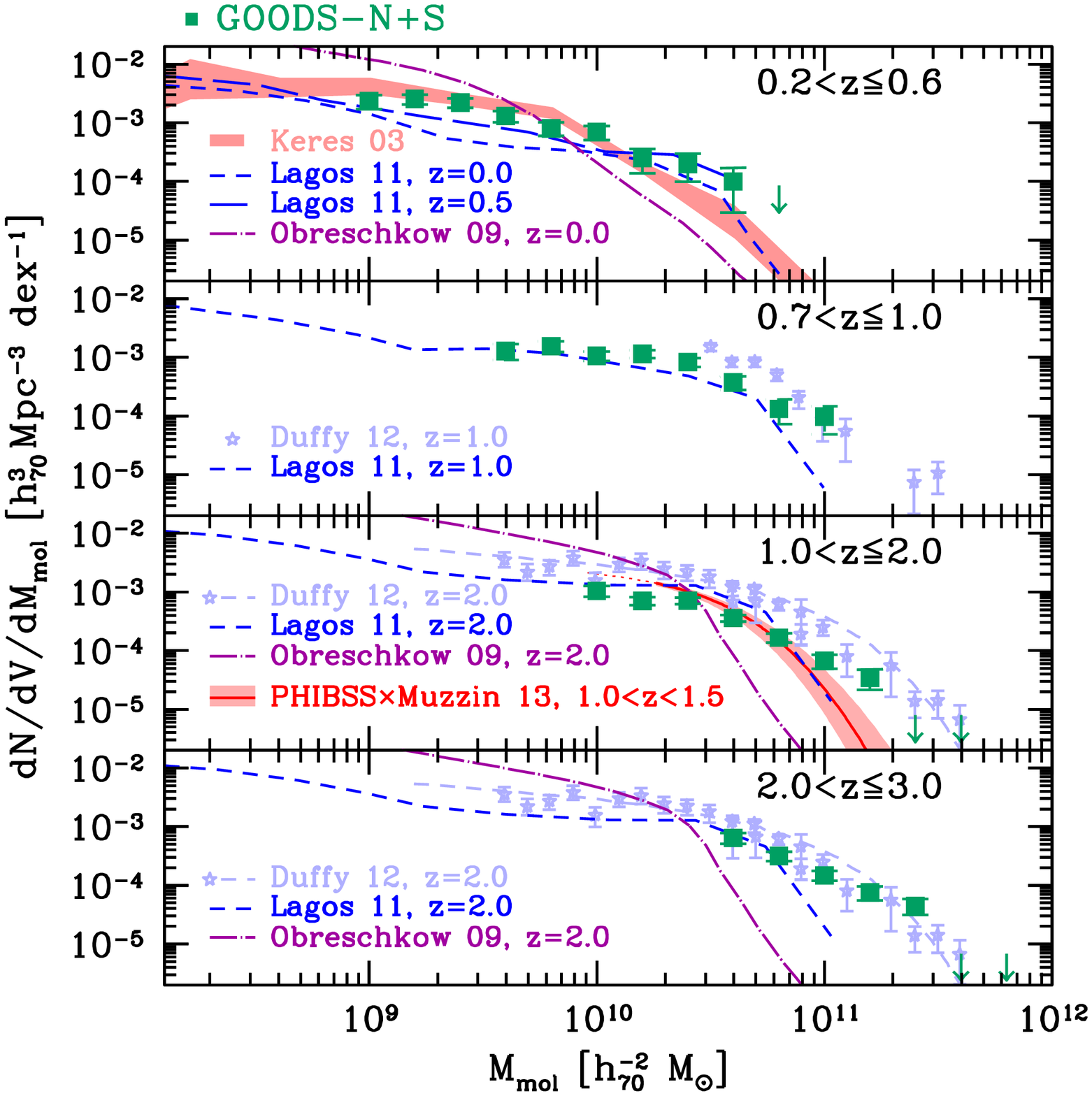}
\includegraphics[width=0.42\textwidth]{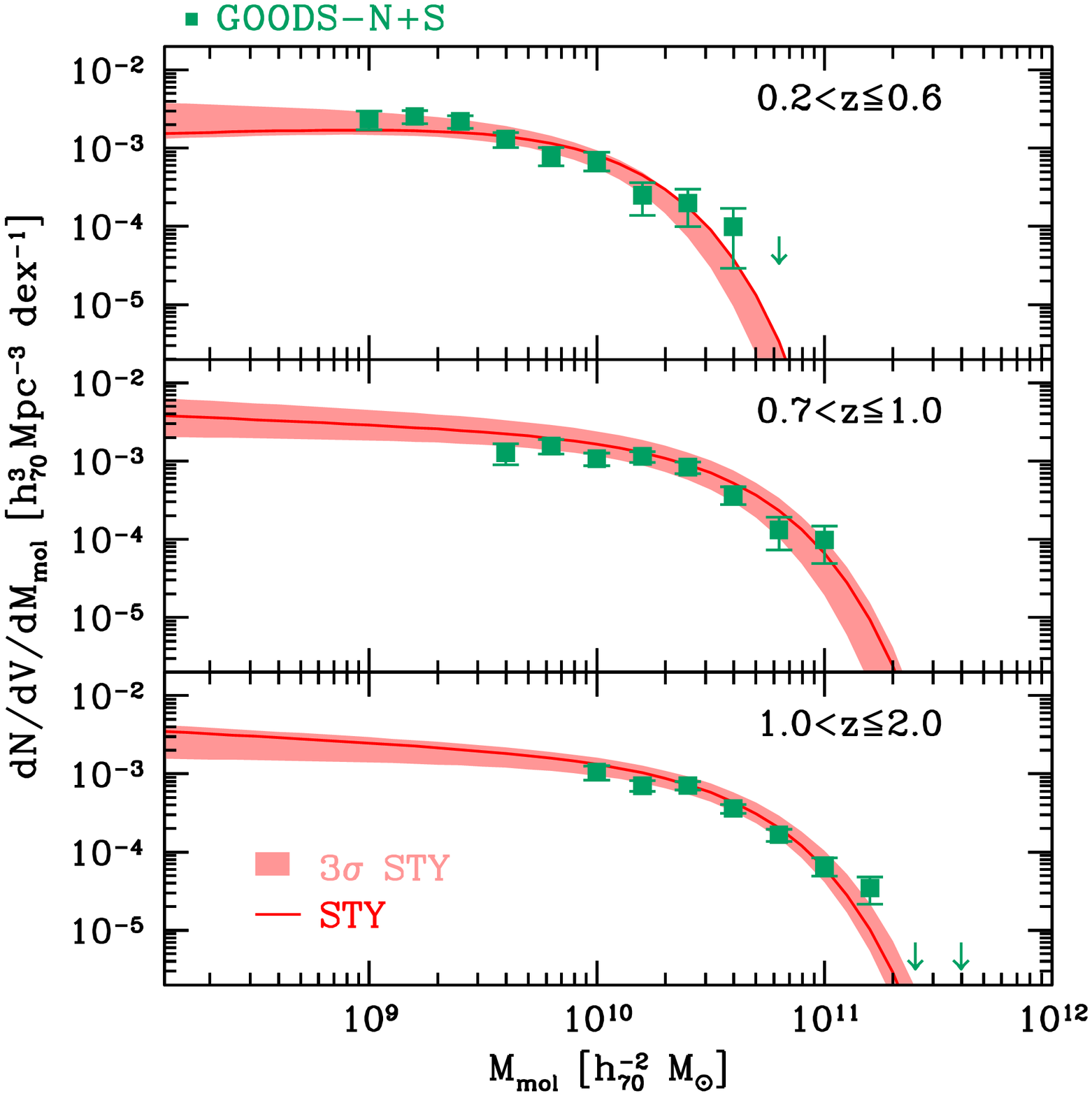}  
\caption{The molecular gas mass function of \textit{Herschel} galaxies. {\em Left}: comparison of 
the $1/V_{\rm a}$ estimate (green squares, based on Eq. \ref{eq:SA_method}) to literature data \citep{keres2003} and models
\citep{obreschkow2009b,lagos2011,duffy2012}. The red line and shaded area at $z=1.0-2.0$ are obtained 
scaling the \citet{muzzin2013} stellar mass function using the 
molecular gas fractions by \citet{tacconi2012}.
When needed, masses found in the literature were scaled by the factor 1.36 necessary to account for Helium, and matched 
to our set of cosmological parameters.
{\em Right}: results of the parametric STY evaluation of a Schechter mass function (red lines) and its 3 
$\sigma$ uncertainty (shaded areas).}
\label{fig:mf}
\end{figure*}

The comoving number density of galaxies in intervals of $M_{\rm mol}$ 
is computed adopting the well known $1/V_{\rm a}$ formalism:
\begin{equation}
\Phi(M)\Delta M=\sum_i \frac{1}{V_{\rm a}^i}\Delta M
\end{equation}
where the sum is computed 
over all sources in the given $M_{\rm mol}$ bin.
The accessible volume $V_{\rm a}$ is a spherical shell delimited by   
$z_{\rm min}^{\rm bin}$ and $\textrm{min}\left(z_{\rm max},z_{\rm max}^{\rm bin}\right)$. Here
$z_{\rm max}$ is the maximum redshift  at which a 
galaxy would be observable in our survey \citep{schmidt1968}, and $z_{\rm min,max}^{\rm bin}$ define 
redshift bins.

Source selection is mainly based on a 160 $\mu$m cut, and the UV requirements do not produce significant 
source losses. On the other hand, the conversion from $S(160)$ to 
$L(\textrm{IR})$ is obtained adopting a family of SED templates spanning a 
variety of colors. Moreover UV properties are also involved in 
computing total SFR and in applying Eq. 1.

As a consequence, the conversion between 160 $\mu$m fluxes and molecular gas mass is not unique,    
but comprises a distribution of ``mass to light ratios''. 
Because of this scatter in $M_{\rm mol}$ vs. $S(160)$, a flux cut induces a 
molecular mass incompleteness. This effect was thoroughly studied by \citet{fontana2004} and \citet{berta2007b}, 
among others, in the case of stellar mass. Equivalently, the recipe defined by these authors can be applied 
to the specific case of molecular gas mass and FIR fluxes to derive completeness corrections as a function of $M_{\rm mol}$.
Using the distribution of the parent MS population, and comparing it to 
that of PACS galaxies leads to similar results.

The comoving number density of galaxies is shown in Fig. \ref{fig:mf}. 
Results from the $\tau_{\rm dep}$ scaling or Eq. \ref{eq:SA_method} are very similar, thus only 
the latter are shown. Table \ref{tab:mf2} reports both estimates, along with 
average completeness values for each mass bin, which also includes the photometric completeness of the FIR catalogs.


An independent characterization of the mass function is provided 
by the maximum likelihood approach by Sandage, Tammann \& Yahil (\citeyear{sandage1979}, STY).
We adopt the Bayesian implementation by \citet{berta2007b} and adapt it to our case,  
thus fully propagating the actual $M_{\rm mol}$ uncertainties into the parametric function evaluation. 
The adopted functional form to describe the mass function is a \citet{schechter1976} function:
\begin{equation}
\Phi(M)dM=\frac{\Phi^\ast}{M^\ast} \left(\frac{M}{M^\ast}\right)^\alpha e^{-\frac{M}{M^\ast}} dM \textrm{,}
\end{equation}
where $\Phi^\ast$ represents the normalization,
$\alpha$ the slope in the low mass regime, and $M^\ast$ the transition
mass between a power-law and the exponential drop-off and the
e-folding mass of the latter. \citet{berta2007b} provide more details about 
this method. Table \ref{tab:mf3} includes the most probable parameter values 
and their 3 $\sigma$ uncertainties, obtained with both $M_{\rm mol}$ estimates. 
The right-hand panel of Fig. \ref{fig:mf} 
compares the result of the STY analysis to the $1/V_{\rm a}$ mass function.
No STY analysis is attempted for the highest redshift bin, because 
only the very massive end is covered.

The Schechter $M^\ast$ parameter increases by 
more than a factor of 3 between $z=0.4$ and 0.8, and then flattens at $z>1$. 
This rate resembles the evolution 
of the IR luminosity function of normal star forming galaxies \citep{gruppioni2013}, 
and reflects the link between SFR and $M_{\rm mol}$.
At the same time, $\Phi^\ast$ varies by only a factor of 2 over the 
0.4-1.5 redshift range. The net effect is a significant evolution of the 
number density of galaxies with large $M_{\rm mol}$, while at the low mass end it remains roughly 
constant. Finally, the most probable value of $\alpha$ 
steepens as redshift increases, but this might be simply an effect of the 
different mass ranges effectively constrained 
at different redshifts (note that the 3 $\sigma$ 
confidence levels are consistent with nearly no evolution).

We compare this $M_{\rm mol}$ mass function to an estimate based on 
stellar mass. Using the PHIBSS CO survey, \citet[][see their Fig. 12]{tacconi2012}
compute molecular gas fractions for $z=1.0-1.5$ normal star forming galaxies, 
as a function of $M_\star$.
We combine these gas fractions with the $z=1.0-1.5$ stellar mass function 
of star forming galaxies by \citet[][see also Ilbert et al. \citeyear{ilbert2013}; Drory et al. \citeyear{drory2009}]{muzzin2013}.
To account for
the MS width in this estimate, we apply a 0.2 dex Gaussian smoothing.
The result is close to the observed mass function at $z=1.0-2.0$ 
(Fig. \ref{fig:mf}). Note also the related approach of 
Sargent et al. (in prep.), as quoted in \citet{carilli2013}.

\section{Discussion}

Our observed $M_{\rm mol}$ function
is computed only for normal star forming galaxies within $\pm0.5$ dex in SFR
from the main sequence. Thus it represents a lower 
limit to the total $M_{\rm mol}$ mass function, by missing passive (low SFR) 
galaxies and powerful, above-sequence objects.
Passive galaxies will provide a negligible contribution 
unless they include                      
a hypothetical population of molecular gas-rich galaxies forming stars with 
very low efficiency. 
For reference, local passive galaxies, lying at $\Delta(\textrm{sSFR})_{\rm MS}\le -1.0$ dex, 
have molecular gas fractions $f_{\rm mol}\le2$\% \citep{saintonge2012}.
Also star bursting galaxies, i.e. 
those lying above the MS, should play only a minor role in shaping the 
mass function. \citet{rodighiero2011} have shown that the contribution of 
above-sequence sources to the number density of star forming galaxies and
total star formation rate density at $z=1.5-2.5$ is small. For 
our $\Delta\log(\textrm{SFR})_{\rm MS}=0.5$ cut, we find that objects above 
the MS contribute no more than $\sim 4$\% to the 
number density of star forming galaxies, and $\sim 16$\% to their SFR density. 
Depending on whether galaxies rise above the main sequence mostly due to an
increased star formation efficiency at fixed gas mass, or due to larger
gas masses, the effect on the mass function will vary between a global upward
$\sim 4$\% shift or a preferential increase at higher gas masses within the 
limits permitted by the 16\% SFR contribution. \citet{saintonge2012} 
show that in local starbursts the two effects share a 50\%-50\% role 
in causing SFR changes with respect to the MS.

\begin{figure}[!ht]
\centering
\includegraphics[width=0.44\textwidth]{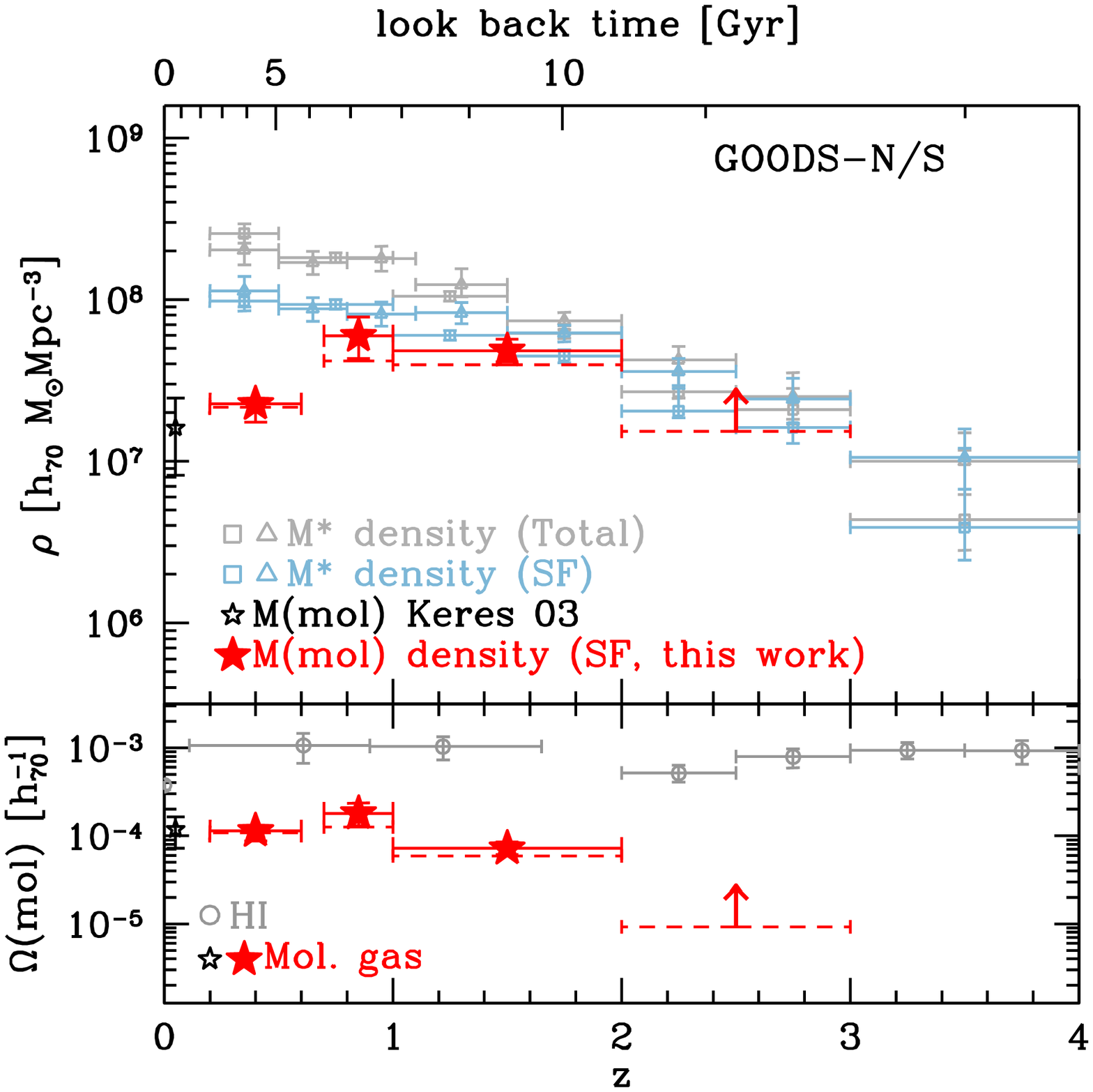}
\caption{{\em Top}: redshift evolution of the molecular gas mass density, based on Eq. \ref{eq:SA_method}. 
Red symbols and solid error bars belong 
to the total mass density; dashed error bars mark lower limits limited to the mass range covered by observations. The black star 
is computed by integrating the local mass function \citep{keres2003}.
Grey squares and triangles represent the total $M_\star$ density by \citet{ilbert2013} and \citet{muzzin2013}, respectively. 
Light blue symbols belong to star forming galaxies only. {\em Bottom}: evolution of 
$\Omega_{\rm mol}$ 
(red symbols, this work) and  $\Omega_{\rm HI}$ \citep{prochaska2004,zwaan2005,rao2006}, where we have 
divided the latter by 1.3 in order to avoid counting Helium twice.}
\label{fig:density}
\end{figure}

Our results are compared in Fig. \ref{fig:mf} to 
the \citet{keres2003} local CO mass function of IRAS-selected galaxies. 
Applying a variable CO-to-H$_2$ conversion factor, inversely dependent on 
$L(\textrm{CO})$ itself, \citet{obreschkow2009a,obreschkow2009b} obtained a revisited mass function quickly dropping at 
the high-mass end. Modeling was implemented by post-processing 
the \citet{delucia2007} semi analytic (SAM) results, 
and assigning the atomic and molecular gas content to galaxies via  a set of physical 
prescriptions \citep[see][]{obreschkow2009c}.
Their expectations (Fig. \ref{fig:mf}) tend to predict a too steep mass function both at 
low and high redshift.
A second model based on the SAM approach was developed by \citet{lagos2011}, starting from 
the \citet{bower2006} galaxy formation model.
As in the case of \citet{obreschkow2009b} the \citet{blitz2006} star formation law is adopted, but 
in this case it is implemented throughout the galaxy evolution process in the SAM model.
The model was tested against the observed stellar mass density evolution and the atomic and molecular
gas content of local galaxies.
Results are shown in Fig. \ref{fig:mf}, and are now much closer to our observed mass function 
up to $z\sim2$. 
%
Figure \ref{fig:mf} finally includes predictions of the hydrodynamical 
simulation by \citet{duffy2012}, 
which overall tend to overestimate the mass function, but are consistent 
with data at the observed high-mass tail.

Adopting the STY Schechter results, we integrate the mass function 
from $10^7$ M$_\odot$ to infinity and obtain a measure of the molecular gas mass density 
(see Table \ref{tab:mf3}
and top panel of Fig. \ref{fig:density}).
A second estimate obtained by limiting the integral to the mass range 
effectively covered by observations is also provided. 
The molecular gas mass density increases by a factor of $\sim4$ from $z=0$ to $z=1$ and then 
remains almost constant up to $z\sim2$. For reference we plot also the $M_\star$ density 
\citep{muzzin2013,ilbert2013} up to $z=4$. The different trends 
reflect the growth of gas fractions as a function of redshift 
\citep[e.g.,][]{tacconi2012}.

Finally, we compute the redshift 
evolution of the density parameter $\Omega_{\rm mol}(z)=\rho_{\rm mol}(z)/\rho_c(z)$ (Fig. \ref{fig:density} 
and Table \ref{tab:mf3}), 
where the critical density at the given redshift is given by $\rho_c(z)=\frac{3 H^2(z)}{8\pi G}$. The molecular gas density parameter peaks at 
$z\sim1.0$; for comparison $\Omega_{\rm HI}$, derived from damped Ly-$\alpha$ systems \citep{prochaska2004,rao2006}, 
and including also low column density cases, does not evolve 
between $z$=$0.5$ and 4.0. 

Using the deepest \textit{Herschel} extragalactic observations available 
and restframe UV information for roughly 700 main sequence galaxies, we have built the 
first molecular gas mass function at redshift $z>0$, and the first estimate 
of its density evolution up to $z=3$. 
While future mm/submm surveys will significantly improve our knowledge of the  
molecular content in high-$z$ galaxies, we have provided a basis for 
refinement of galaxy evolution models that are accounting for the molecular 
phase.


\begin{acknowledgements}  
PACS has been developed by a consortium of institutes led by MPE (Germany) and 
including UVIE (Austria); KU Leuven, CSL, IMEC (Belgium); CEA, LAM (France); 
MPIA (Germany); INAF-IFSI/OAA/OAP/OAT, LENS, SISSA (Italy); IAC (Spain). 
This development has been supported by the funding agencies BMVIT (Austria), 
ESA-PRODEX (Belgium), CEA/CNES (France), DLR (Germany), ASI/INAF (Italy), 
and CICYT/MCYT (Spain). 
\end{acknowledgements}




\bibliographystyle{aa}
\bibliography{biblio_PEP_Mmol_MF}


\begin{landscape}

\begin{table*}[!ht]
\caption{The molecular gas mass function of \textit{Herschel} galaxies, derived with the $1/V_{\rm a}$ method.}
\scriptsize
\centering
\begin{tabular}{c | c c c | c c c | c c c | c c c }
\hline
\hline
                                 & \multicolumn{3}{c|}{$0.2<z\le0.6$} & \multicolumn{3}{c|}{$0.7<z\le1.0$} & \multicolumn{3}{c|}{$1.0<z\le2.0$} & \multicolumn{3}{c}{$2.0<z\le3.0$} \\
$\log(M_{\rm mol})$       & $\Phi(\textrm{M})$ & $\Phi(\textrm{M})$ & Compl. & $\Phi(\textrm{M})$ & $\Phi(\textrm{M})$ & Compl. & $\Phi(\textrm{M})$ & $\Phi(\textrm{M})$ & Compl. & $\Phi(\textrm{M})$ & $\Phi(\textrm{M})$ & Compl. \\
			  & Eq. \ref{eq:SA_method}		& $\tau_{\rm dep}$		    &  & Eq. \ref{eq:SA_method}		& $\tau_{\rm dep}$		    &  & Eq. \ref{eq:SA_method}		& $\tau_{\rm dep}$		    &  & Eq. \ref{eq:SA_method}		& $\tau_{\rm dep}$		    &  \\
$[h_{70}^{-2}\ \textrm{M}_\odot]$ & \multicolumn{2}{c}{$[10^{-4}\ h_{70}^3\ \textrm{Mpc}^{-3} \textrm{dex}^{-1}]$}  &  & \multicolumn{2}{c}{$[10^{-4}\ h_{70}^3\ \textrm{Mpc}^{-3} \textrm{dex}^{-1}]$} &  & \multicolumn{2}{c}{$[10^{-4}\ h_{70}^3\ \textrm{Mpc}^{-3} \textrm{dex}^{-1}]$} &  & \multicolumn{2}{c}{$[10^{-4}\ h_{70}^3\ \textrm{Mpc}^{-3} \textrm{dex}^{-1}]$} & \\
\hline
    9.0   &  23.42$\pm$6.26	& -- 			&   0.52  &	  --		 & --			&   -- & --  	  	& --		&  --  & -- 	   	& --		& -- \\
    9.2   &  25.38$\pm$4.80	& 18.42$\pm$5.11	&   0.67  &	  --		 & --			&   -- & --  	  	& --		&  --  & -- 	   	& --		& -- \\
    9.4   &  22.01$\pm$4.02	& 24.42$\pm$4.39	&   0.78  &	  --		 & --			&   -- & --  	  	& --		&  --  & -- 	   	& --		& -- \\
    9.6   &  13.00$\pm$2.77	& 17.86$\pm$3.32	&   0.86  &	12.84$\pm$3.87   & 20.15$\pm$6.08	& 0.61 & --  	  	& --		&  --  & -- 	   	& --		& -- \\
    9.8   &   8.06$\pm$2.08	& 16.85$\pm$2.98	&   0.98  &	15.54$\pm$3.24   & 11.50$\pm$2.64	& 0.75 & --  	  	& --		&  --  & -- 	   	& --		& -- \\
   10.0   &   6.97$\pm$1.86	& 10.46$\pm$2.28	&   1.00  &	10.67$\pm$1.98   & 13.53$\pm$2.02	& 0.85 & 10.47$\pm$2.18	& 15.47$\pm$2.82& 0.57 & -- 	   	& --		& -- \\
   10.2   &   2.49$\pm$1.11	& 4.48$\pm$1.49		&   1.00  &	11.43$\pm$1.79   & 10.52$\pm$1.64	& 0.96 & 7.05$\pm$1.09 	& 7.27$\pm$1.01	& 0.72 & -- 	   	& --		& -- \\
   10.4   &   1.99$\pm$1.00	& 1.99$\pm$1.00		&   1.00  &	 8.33$\pm$1.43   & 6.37$\pm$1.25	& 1.00 & 7.09$\pm$0.85 	& 5.37$\pm$0.65	& 0.82 & -- 	   	& --		& -- \\
   10.6   &   1.00$\pm$0.70	& 0.50$\pm$0.50 	&   1.00  &	 3.74$\pm$0.97   & 4.17$\pm$1.01	& 1.00 & 3.60$\pm$0.47 	& 2.65$\pm$0.38	& 0.90 & 6.41$\pm$1.34 	& 6.47$\pm$1.08	& 0.61\\
   10.8   &   0.50$\pm$0.50	& --			&   1.00  &	 1.32$\pm$0.59   & 1.22$\pm$0.55 	& 1.00 & 1.67$\pm$0.30 	& 1.64$\pm$0.28	& 1.00 & 3.15$\pm$0.57 	& 1.84$\pm$0.36	& 0.76\\
   11.0   &    -- 		& --			&    --   &	 0.98$\pm$0.49   & --			& 1.00 & 0.67$\pm$0.18 	& 0.38$\pm$0.13	& 1.00 & 1.48$\pm$0.28 	& 1.19$\pm$0.24	& 0.86\\
   11.2   &    -- 		& --			&    --   &	  --		 & --			&   -- & 0.35$\pm$0.13 	& 0.28$\pm$0.12	& 1.00 & 0.75$\pm$0.20 	& 0.81$\pm$0.20	& 0.94\\
   11.4   &    -- 		& --			&    --   &	  --		 & --			&   -- & 0.05$\pm$0.05 	& 0.05$\pm$0.05	& 1.00 & 0.45$\pm$0.13 	& 0.20$\pm$0.09	& 1.00\\
   11.6   &    -- 		& --			&    --   &	  --		 & --			&   -- & 0.05$\pm$0.05 	& --		& 1.00 & 0.04$\pm$0.04 	& 0.08$\pm$0.06	& 1.00\\
   11.8   &    -- 		& --			&    --   &	  --		 & --			&   -- & --  	  	& --		&  --  & 0.04$\pm$0.04 	& --		& 1.00\\
\hline
Tot. Num. & \multicolumn{3}{c|}{145} & \multicolumn{3}{c|}{166} & \multicolumn{3}{c|}{260} & \multicolumn{3}{c}{122} \\
\hline
\end{tabular}
\normalsize
\label{tab:mf2}
\end{table*}

\begin{table*}[!ht]
\caption{Results of STY analysis, and molecular gas mass density. {\em Top}: results based on Eq. \ref{eq:SA_method}. 
{\em Bottom}: results obtained with the $\tau_{\rm dep}$ scaling of SFR.}
\scriptsize
\centering
\begin{tabular}{l | c c c | c c c | c c c | c c c }
\hline
\hline
				& \multicolumn{3}{c|}{$0.2<z\le0.6$} & \multicolumn{3}{c|}{$0.7<z\le1.0$} & \multicolumn{3}{c|}{$1.0<z\le2.0$} & \multicolumn{3}{c}{$2.0<z\le3.0$} \\
Based on Eq. \ref{eq:SA_method}	& Most prob.	& Min.  & Max.  &   Most prob.	& Min.  & Max.  &  Most prob.	& Min.  & Max.  &  Most prob.	& Min.  & Max.  \\ 
\hline
$\Phi^\ast$    $[10^{-4}\ h_{70}^3\ \textrm{Mpc}^{-3} \textrm{dex}^{-1}]$ & 11.0  &  9.8  & 11.0  & 7.8   &  7.4  & 10.6  &  5.6  & 5.3   & 6.6   & -- & -- & -- \\
$\log(M^\ast)$ $[h_{70}^{-2}\ \textrm{M}_\odot]$                          & 10.34 & 10.23 & 10.34 & 10.84 & 10.72 & 10.86 & 10.89 & 10.84 & 10.94 & -- & -- & -- \\
$\alpha$                                                                  & -0.91 & -1.08 & -0.90 & -1.12 & -1.15 & -1.03 & -1.16 & -1.16 & -1.04 & -- & -- & -- \\
\hline
$\rho_{\rm mol}(\textrm{tot})$ $[10^7\ h_{70}\ \textrm{M}_\odot\ \textrm{Mpc}^{-3}]$& 2.266 & 1.751 & 2.289 & 5.974 & 4.317 & 7.818 & 4.836 & 4.112 & 5.711 & -- & -- & -- \\
$\rho_{\rm mol}(\textrm{lim})$ $[10^7\ h_{70}\ \textrm{M}_\odot\ \textrm{Mpc}^{-3}]$& 2.166 & 1.619 & 2.188 & 4.173 & 3.231 & 5.361 & 3.958 & 3.301 & 4.972 & 1.533 & 1.184 & 1.882 \\
\hline
$\Omega_{\rm mol}(\textrm{tot})$ $[10^{-5}\ h_{70}^{-1}]$  & 11.34 & 8.76 & 11.46 & 18.00 & 13.01 & 23.56 & 7.19 & 6.12 & 8.50 & -- & -- & -- \\
$\Omega_{\rm mol}(\textrm{lim})$ $[10^{-5}\ h_{70}^{-1}]$  & 10.84 & 8.10 & 10.95 & 12.57 & 9.74 & 16.15 & 5.89 & 4.91 & 7.40 & 0.92 & 0.71 & 1.13 \\
\hline
\multicolumn{13}{c}{}\\
\hline
				& \multicolumn{3}{c|}{$0.2<z\le0.6$} & \multicolumn{3}{c|}{$0.7<z\le1.0$} & \multicolumn{3}{c|}{$1.0<z\le2.0$} & \multicolumn{3}{c}{$2.0<z\le3.0$} \\
Based on $\tau_{\rm dep}$	& Most prob.	& Min.  & Max.  &   Most prob.	& Min.  & Max.  &  Most prob.	& Min.  & Max.  &  Most prob.	& Min.  & Max.  \\ 
\hline
$\Phi^\ast$    $[10^{-4}\ h_{70}^3\ \textrm{Mpc}^{-3} \textrm{dex}^{-1}]$ & 12.0  &  9.8  & 13.6  & 7.4   &  6.8  & 9.3  &  4.9  & 4.3   & 6.2   & -- & -- & -- \\
$\log(M^\ast)$ $[h_{70}^{-2}\ \textrm{M}_\odot]$                          & 10.32 & 10.27 & 10.36 & 10.79 & 10.70 & 10.83 & 10.87 & 10.83 & 10.93 & -- & -- & -- \\
$\alpha$                                                                  & -1.01 & -1.11 & -1.00 & -1.06 & -1.10 & -1.00 & -1.11 & -1.14 & -1.04 & -- & -- & -- \\
\hline
$\rho_{\rm mol}(\textrm{tot})$ $[10^7\ h_{70}\ \textrm{M}_\odot\ \textrm{Mpc}^{-3}]$& 2.540 & 1.963 & 3.114 & 4.725 & 3.640 & 6.287 & 3.880 & 3.205 & 5.407 & -- & -- & -- \\
$\rho_{\rm mol}(\textrm{lim})$ $[10^7\ h_{70}\ \textrm{M}_\odot\ \textrm{Mpc}^{-3}]$& 2.400 & 1.810 & 2.943 & 3.437 & 2.771 & 4.407 & 3.232 & 2.590 & 4.698 & 1.213 & 0.929 & 1.497 \\
\hline
$\Omega_{\rm mol}(\textrm{tot})$ $[10^{-5}\ h_{70}^{-1}]$  & 12.71 & 9.82 & 15.59 & 14.21 & 10.97 & 18.94 & 5.77 & 4.77 & 8.04 & -- & -- & -- \\
$\Omega_{\rm mol}(\textrm{lim})$ $[10^{-5}\ h_{70}^{-1}]$  & 12.01 & 9.06 & 14.73 & 10.36 & 8.35 & 13.28 & 4.81 & 3.85 & 6.99 & 0.73 & 0.56 & 0.90 \\
\hline
\end{tabular}
\tablefoot{Minimum and maximum values are computed at the 3 $\sigma$ confidence level. The molecular gas mass density $\rho(\textrm{tot})$ is obtained by integrating the mass function between $10^7$ M$_\odot$ and infinity; 
$\rho(\textrm{lim})$ is computed solely on the mass range covered by PEP data (see Table \ref{tab:mf2}).}
\normalsize
\label{tab:mf3}
\end{table*}

\end{landscape}






\end{document}